\documentclass[twocolumn,multicol]{aastex62}
\usepackage{gensymb}

\newcommand{\xtejl}{XTE~J1810$-$197}
\newcommand{\xtej}{J1810}
\newcommand{\nustar}{{\it NuSTAR}}
\newcommand{\nicer}{{\it NICER}}
\newcommand{\swift}{{\it Swift}}

\graphicspath{{./}{figures/}}

\received{January 1, 2019}
\revised{January 7, 2019}
\accepted{\today}
\submitjournal{ApJL}

\shorttitle{{\it NICER} observations of XTE~J1810$-$197}
\shortauthors{G\"uver et al.}
\begin{document}

\title{{\it NICER} Observations of the 2018 Outburst of \xtejl}

\correspondingauthor{Tolga G\"uver}
\email{tolga.guver@istanbul.edu.tr}

\author[0000-0002-3531-9842]{Tolga G\"uver}
\affil{Department of Astronomy and Space Sciences, Faculty of Science, \.{I}stanbul University, 34119, \.{I}stanbul, Turkey}
\affil{\.{I}stanbul University Observatory Research and Application Center, \.{I}stanbul University 34119, \.{I}stanbul Turkey}
\author{Ersin G{\"o}{\u g}{\"u}{\c s}}
\affil{Faculty of Engineering and Natural Sciences, Sabanc\i University, Orhanl\i-Tuzla 34956, \.{I}stanbul, Turkey }
\author{Eda Vurgun}
\affil{\.{I}stanbul University, Graduate School of Sciences, Department of Astronomy and Space Sciences, 34116, Beyaz\i t, \.{I}stanbul, Turkey }
\author[0000-0003-1244-3100]{Teruaki Enoto}
\affiliation{Department of Astronomy, Kyoto University, Kitashirakawa-Oiwake-cho, Sakyo-ku, Kyoto 606-8502, Japan}
\author{Keith C. Gendreau}
\affiliation{X-ray Astrophysics Laboratory, Astrophysics Science Division, NASA's Goddard Space Flight Center, Greenbelt, MD 20771, USA}
\author{Takanori Sakamoto}
\affil{Department of Physics and Mathematics, Aoyama Gakuin University, 5-10-1 Fuchinobe, Chuo-ku, Sagamihara-shi Kanagawa 252-5258, Japan}
\author[0000-0003-3847-3957]{Eric V. Gotthelf}
\affil{Columbia Astrophysics Laboratory, Columbia University, 550 West 120th Street, New York, NY 10027, USA}
\author{Zaven Arzoumanian}
\affiliation{X-ray Astrophysics Laboratory, Astrophysics Science Division, NASA's Goddard Space Flight Center, Greenbelt, MD 20771, USA}
\author[0000-0002-6449-106X]{Sebastien Guillot} \affil{CNRS, IRAP, 9 avenue du Colonel Roche, BP 44346, F-31028 Toulouse Cedex 4, France}
\author[0000-0002-6789-2723]{Gaurava K. Jaisawal}
\affil{National Space Institute, Technical University of Denmark, 
  Elektrovej 327-328, DK-2800 Lyngby, Denmark}
\author[0000-0002-0380-0041]{Christian Malacaria}
\affiliation{NASA Marshall Space Flight Center, NSSTC, 320 Sparkman Drive, Huntsville, AL 35805, USA}\thanks{NASA Postdoctoral Fellow}
\affiliation{Universities Space Research Association, NSSTC, 320 Sparkman Drive, Huntsville, AL 35805, USA}
\author[0000-0002-4694-4221]{Walid A. Majid}
\affiliation{Jet Propulsion Laboratory, California Institute of Technology, Pasadena, CA 91109, USA}
\affiliation{Division of Physics, Mathematics, and Astronomy, California Institute of Technology, Pasadena, CA 91125, USA}



\begin{abstract}

We present the earliest available soft X-ray observations of \xtejl, the prototypical
transient magnetar, obtained 75--84 days after its 2018 outburst with the \textit{Neutron Star Interior Composition Explorer} (\nicer).  Using a series of observations covering eight days we find that its decreasing X-ray flux is well-described by either a blackbody plus power-law or a two-blackbody spectral model.
The 2--10~keV flux of the source varied from $(1.206\pm0.007)\times10^{-10}$ to $(1.125\pm0.004)\times10^{-10}~{\rm erg}~{\rm s}^{-1}~{\rm cm}^{-2}$, a decrease of about 7\% within our observations and 44\% from that measured 7--14~days after the outburst with \textit{NuSTAR}. We confirm that the pulsed fraction and spin pulse phase of the neutron star are energy dependent up to at least 8~keV. Phase resolved spectroscopy of the pulsar suggests magnetospheric variations relative to the line of sight.

\end{abstract}



\section{Introduction} 
\label{sec:intro}

Anomalous X-ray Pulsars (AXPs) and Soft Gamma-ray Repeaters (SGRs) are thought to be the observational manifestations of strongly magnetized neutron stars, magnetars. AXPs and SGRs differentiate themselves from other isolated pulsars with their relatively long spin periods in the range of 2--12~s and large spin-down rates, between $10^{-13}$ and $10^{-10}$~s~s$^{-1}$, implying a dipole magnetic field strength that is almost two orders of magnitude larger than that of most radio pulsars \citep{Kaspi2017,Olausen2014}. The magnetar model \citep{Duncan1992, Thompson1996}, naturally explains most of their observational properties, assuming that these objects indeed have strong dipole and/or internal magnetic fields in the range $B\approx10^{14-15}$~G. 

Historically, magnetars have been recognized by their bright persistent X-ray luminosities in the range of 10$^{33-36}$~erg~s$^{-1}$. However, observations in the last two decades have revealed a growing number of transient magnetars. These sources show a sudden increase in X-ray brightness followed by a slower decay, with typical timescales of months to years \citep{Zelati2018}. The transient AXP \xtejl\ (hereafter \xtej) is the first identified and prototypical transient magnetar. It was discovered in early 2003 by \citep{Ibrahim2004} using the Rossi X-Ray Timing Explorer ({\it RXTE}), when its X-ray luminosity increased by a factor of almost 150, compared to its quiescent state recorded earlier, serendipitously, in Einstein, ROSAT, and ASCA observations \citep{Gotthelf2004}.

\xtej\ has a spin period of 5.54~s and a large period derivative, $\dot{P} \approx 10^{-11}$~s~s$^{-1}$, implying a dipole field  of $3\times10^{14}$~G and a characteristic age $\tau \approx 11000~yr$ \citep{Camilo2007}. The distance to the source is estimated to be $3.1\pm0.5$~kpc \citep{Durant2006}. Archival X-ray observations prior to its original 2003 outburst indicate a soft thermal quiescent spectrum of $kT \approx 0.19$~keV with a
flux of $\approx$7$\times10^{-13}~{\rm erg}~{\rm s}^{-1}~\rm{cm^{-2}}$ (see Table 3, \cite{Gotthelf2004}). The spectral evolution and long-term flux variability have been studied in numerous investigations: Its X-ray spectra have generally been modeled by two blackbody components \citep[e.g.,][]{Gotthelf2004, Gotthelf2007, Guver2007, Bernardini2009, Bernardini2011, Alford2016, Vurgun2019}. These components have been attributed to emission from the whole surface of the neutron star and a hot spot on the surface. 

After more than 15 years, \xtej\  exhibited in 2018 December a very significant flux increase in the radio band (at 1.53 GHz; \citealt{Lyne2018}). The source was then also detected at higher and lower frequencies \citep{Desvignes2018, Lower2018, Majid2019}. Unfortunately, soft X-ray observations could not start immediately due to its position near the Sun at that time. Nevertheless, \textit{MAXI} \citep{Mihara2018} and \nustar\ were able to observe the source \citep{Gotthelf2019}. The peak flux measured in the 2$-$10 keV band with \textit{MAXI} was 2.9$\times$10$^{-10}~{\rm erg}~{\rm s}^{-1}~{\rm cm}^{-2}$. \nustar\ data showed that the X-ray spectrum is similar to that found in 2003 \citep{Ibrahim2004}, with a blackbody temperature of $kT = 0.74$~keV and a non-thermal power-law with an index of 4.4 \citep{Gotthelf2019}. They reported a 2--10 keV flux of $2\times10^{-10}~{\rm erg}~{\rm s}^{-1}~{\rm cm}^{-2}$, which is a factor of 2 greater than the 2003 projected maximum outburst flux \citep{Gotthelf2007}. \citet{Gotthelf2019} also reports an energy dependent modulation and phase shift in the \nustar\ band.

In this letter, we report on follow-up 0.7--8.0 keV X-ray observations
of the unique transient magnetar, taken 2.5 months after the onset of its 2018 outburst. In \S~\ref{sec:observations}, we outline the NICER observations of \xtej, pre and post its 2018 outburst. In \S~\ref{sec:timing}, \S~\ref{sec:sp_analysis} and \S~\ref{sec:phase}, we present the temporal and spectral analyses, including phase-resolved studies. We compare X-ray spectra and pulse profiles obtained before and after the outburst. Finally, we discuss these results in Section \S~\ref{sec:disc}.

\section{{\it NICER} Observations of \xtejl}
\label{sec:observations}

Since its launch in 2017, \nicer\ provides fast-timing spectroscopy observations with a large effective area in soft X-rays (0.2--12~keV) as an external payload on the International Space Station \citep{Gendreau2012}. Shortly after the start of the mission, \nicer\ observed \xtej\ as part of a science team investigation between 2017 August and  2018 July, with observation IDs 0020420104--0020420112 and 1020420101--1020420129. The total exposure time of these observations was 33.4~ks. Throughout the text we refer to these observations as the pre-outburst data. Triggered by reports of the significant radio flux increase, Target of Opportunity (ToO) observations with \nicer\ were performed throughout February 2019 \citep{Guver2019}: data were obtained between 2019 February 6 (MJD 58520.99) and 15 (58529.50), with observation IDs 1020420130--1020420137. We refer to these observations as the post-outburst data.

For calibration and filtering of the data, we used HEASOFT version 6.23 and {\it NICERDAS} version \textit{2018-03-01\_V003}. We applied the standard filtering criteria (excluding events acquired during times of South Atlantic Anomaly passage and with pointing offsets greater than 54\arcsec; including data obtained with Earth elevation angles greater than 30\degree above the dark limb and 45\degree above the bright limb). We also omitted data from detectors \#14, \#34 and \#54, which occasionally register higher electronic noise. We extracted the pre-outburst spectrum from the 2017-2018 observations, imposing a further constraint of Sun angle greater than $60\degree$ to minimize background at soft energies. The resulting exposure time is 27.8~ks. For the post-outburst observations, we had to relax the Sun angle criterion to be greater than $45\degree$, resulting in relatively higher background that restricted our X-ray spectroscopy to the 0.7--8.0~keV range. The final exposure time of the filtered post-outburst data is 14.7~ks.

Because \nicer\ consists of non-imaging detectors, the \nicer\ team has developed a space weather-based background model, which estimates spectral contributions from the time-dependent particle background, optical loading from the Sun, and diffuse sky background using a library of observations of ``blank sky'' fields (Gendreau et al., in prep.). Using this model, post-outburst we obtained an average background-subtracted source rate of 48 counts~s$^{-1}$ 
(0.5 count~s$^{-1}$ background) in the 0.7--8.0~keV band, and for the pre-outburst observations, a source rate of 0.74 counts~s$^{-1}$ (0.28 count~s$^{-1}$ background) in the 0.7--2.5~keV band.

%%%%%%%%%%%%%%%%%%%%%%%%% Temporal Analysis %%%%%%%%%%%%%%%%%%%%%%%%%%%%

\section{Temporal Analysis}
\label{sec:timing}
%% count rate 

 We applied barycentric correction to the 14.7~ks post-outburst data using the source coordinates (J2000) R.A. 18h09m51s and Decl. $-19^{\circ}43^{\prime}51^{\prime\prime}$ \citep{Helfand2007}. We employed the Z$^{2}$ test \citep{Buccheri1983} with one harmonic to search for a periodic signal in the observation on MJD 58522.6 (the epoch), and constructed a template pulse profile. We then calculated the phase shift between pulse profiles of each pointing and the template. Modeling the phase shift trend in time with a second order polynomial, we obtained a phase connected timing solution covering the time span MJD 58521.043--58529.490, yielding a pulse period $P=5.5414819(4)$~s and period derivative  $\dot{P}$=$(9.1\pm1.5)\times10^{-12}$~s~s$^{-1}$. The measured spin frequency is slightly larger ($\Delta P$=4$\times10^{-6}$~s) than the extrapolation of the timing solution provided by \cite{Pintore2019}. The period derivative is consistent with that obtained from contemporaneous radio observations \citep{Levin2019}. 

\begin{figure}[ht!]
\begin{center}
\includegraphics[scale=0.30]{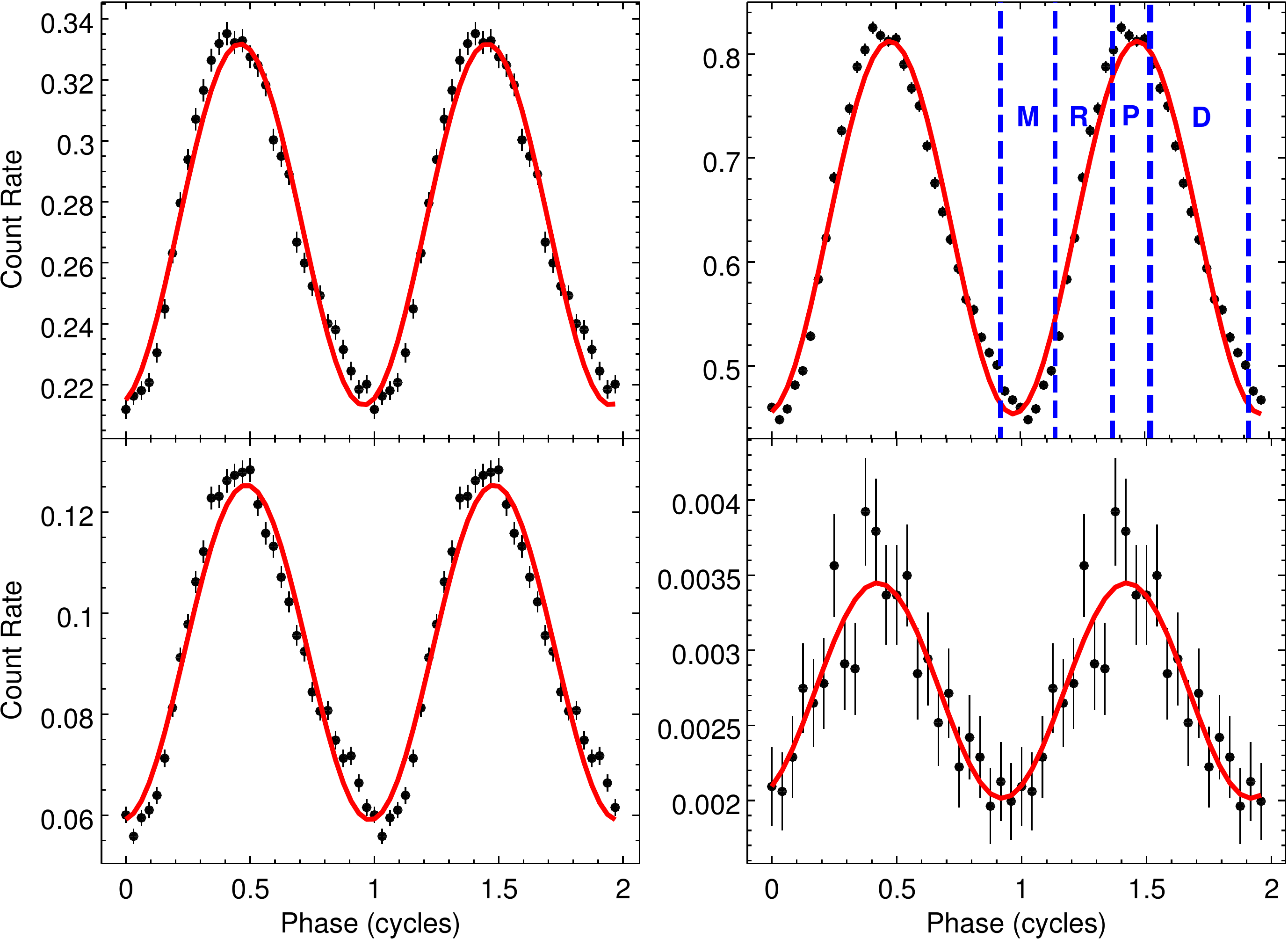}
\caption{Energy dependent X-ray pulse profiles, including background. From left to right and top to bottom, panels show the pulse profiles in the 0.7$-$1.5, 1.5$-$3.5, 3.5$-$6.5, and 6.5$-$8.0~keV bands. Best-fit sinusoidal functions are over-plotted for each pulse profile. Vertical lines in the upper right panel indicate the spin phase intervals selected for phase-resolved spectral analysis. Minimum, Rise, Peak and Decay intervals are labeled M, R, P, and D, respectively.}
\label{pf}
\end{center}
\end{figure}

We have constructed energy resolved pulse profiles using our spin ephemeris. In Figure~\ref{pf}, we present the profiles in the 0.7--1.5 keV, 1.5--3.5 keV, 3.5--6.5 keV, and 6.5--8.0 keV bands, chosen to indicate energy ranges dominated by different spectral components. The pulse profiles are well approximated by a sinusoidal model, with pulse fractions\footnote{The pulsed fraction $f_p$ is defined as the ratio of the pulsed component to the total flux in the pulse profile and is determined here from the fitted sine curves, $f_p =A/B$, where $y(\phi) = A \sin \phi + B$} in the above energy bands of 22.6$\pm$0.2\%, 29.4$\pm$0.2\%, 36.5$\pm$0.4\%, 23.8$\pm$2.1\%, respectively. Similar to the results of \citet{Gotthelf2019}, the pulse fraction is found to increase with energy, up to 6.5~keV, before decreasing again. Note that the pulsed fractions are not background subtracted. With respect to the profile in the lowest energy band, the relative pulse phase in the above mentioned energy ranges are $\phi_{off}$ = $0.000\pm0.003$, $+0.015\pm0.001$, $+0.024\pm0.003$, and $-0.037\pm0.023$,  respectively. This is consistent with the hardest X-ray band pulses arrive faster than in the lower energy bands. Beyond 8.0 keV, background begins to dominate and hence the pulses can not be detected. For comparison with the pre-outburst behavior of the source, we analyzed a 3~ks {\it NICER} observation of \xtej\ performed on 2017 August 27 (ID 1020420114). We detected a weak signal in the 0.7-1.5 keV band. The  pulsed fraction in this energy band is 29.5$\pm$6.2\%, while all other energy bands are consistent with random fluctuations due to background. 

We also performed a search for short bursts as follows: we constructed a light curve with 0.1~s time resolution using X-ray events in the 0.7--12 keV band. Using Poisson statistics, we have determined the probability of rates in every 0.2~s long data (i.e., two consecutive bins) exceeding the mean rate at the 99.99\% confidence level. Note that the mean rate is determined from a time segment either prior to or following the bins under investigation, with a 2~s gap in between. We detect no significant short magnetar-like burst during any \nicer\ data sets on timescales of 0.1~s or longer.

%%%%%%%%%%%%%%%%%%%%%%%% Spectral analysis %%%%%%%%%%%%%%%%%%%%%%%%%%%%

\section{Spectral analysis}
\label{sec:sp_analysis}

\begin{figure}[ht!]
\begin{center}
\includegraphics[scale=0.45,angle=0]{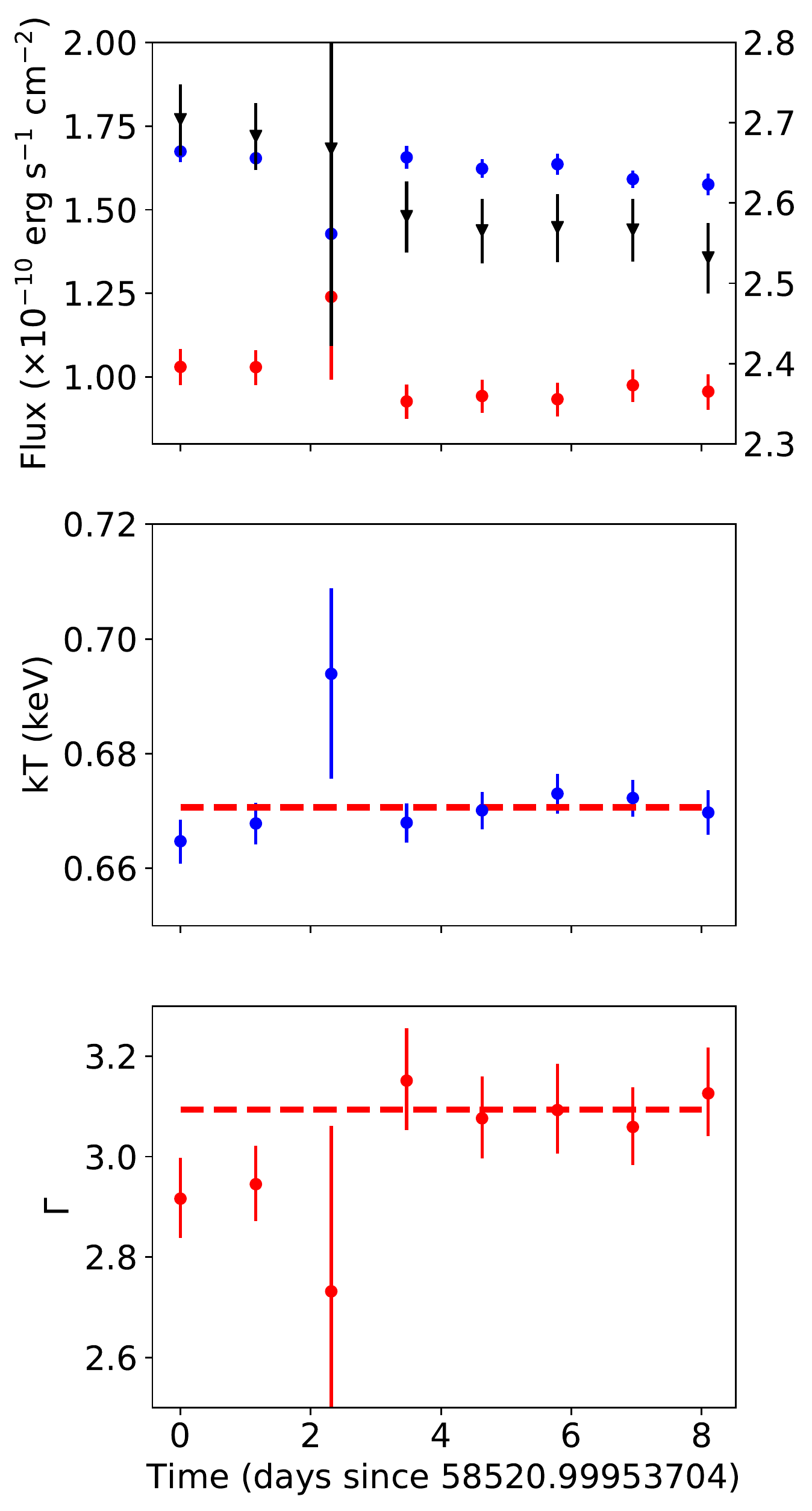}
\caption{ (upper panel) Evolution of unabsorbed total flux (black) as well as that of the blackbody component (blue) and the power law (red) are shown. Note that the scale on the left corresponds to the fluxes of each component while the one on the right is for the total flux. (middle panel) Evolution of the blackbody temperature is shown. The red dashed line shows the weighted average of the last five observations. (bottom panel) Evolution of the power law photon index, and the weighted mean of the last five values in the dashed line are shown.\label{evolution}}
\end{center}
\end{figure}

\begin{figure*}[ht!]
\begin{center}
\includegraphics[scale=0.7,angle=0]{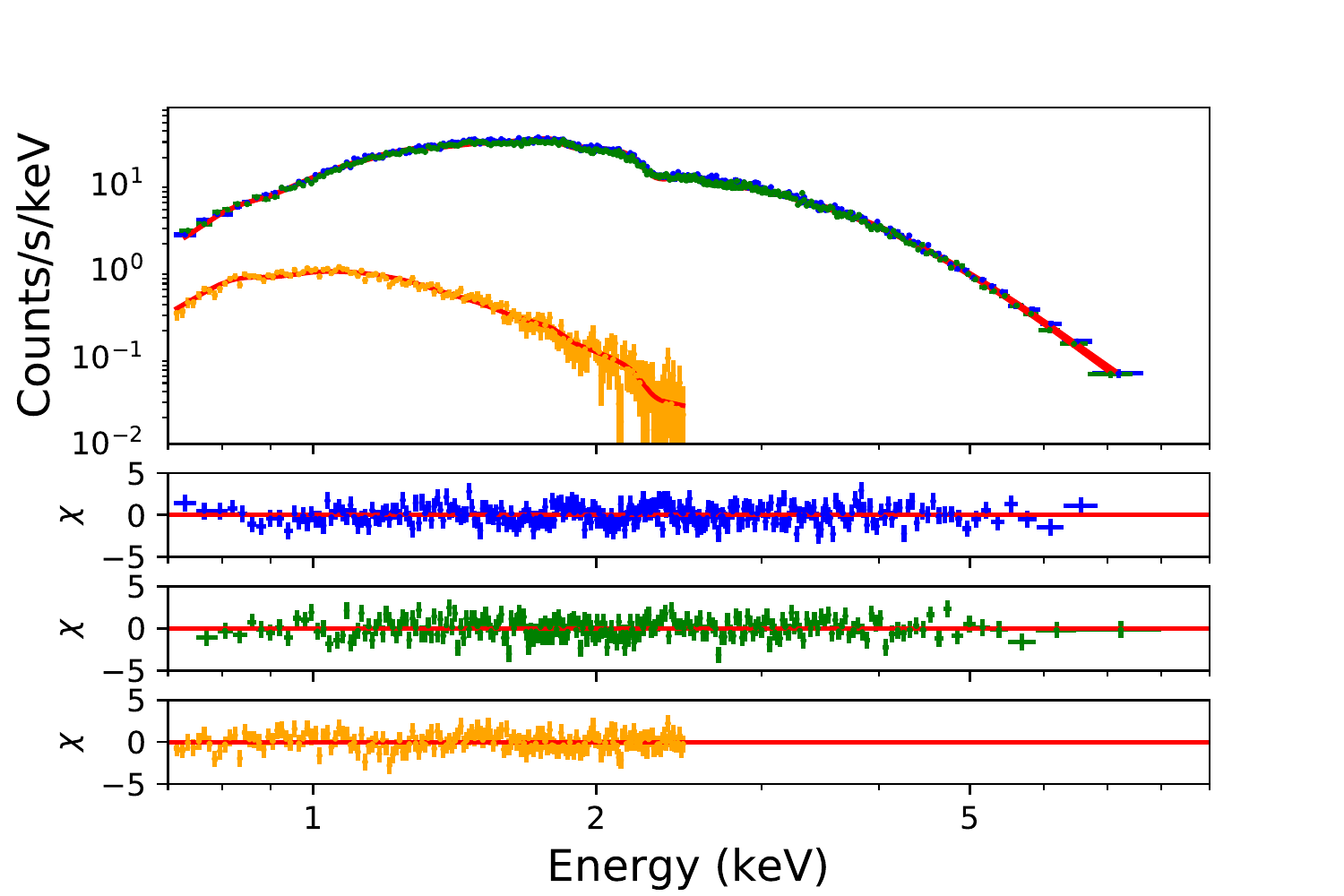}
\caption{{\it NICER} soft-X-ray spectra of \xtej\ obtained throughout 2017--2018 (in orange) as well as in February 2019. For clarity, we show two of the longest exposures (2247s and 2645s, second and seventh segments) obtained in February 2019, one from the beginning and another one from the end of our coverage, with blue and green, respectively. The best fit blackbody plus power-law models are also overplotted and shown in red. The lower three panels show the residuals from the model for these segments, and the pre-outburst data, respectively. \label{spectra}}
\end{center}
\end{figure*}

To search for any spectral evolution within the 8 day span of the post-outburst \nicer\ data set, we analyzed spectra from the daily 2~ks observations. We also included the spectrum of the pre-outburst data in a joint fit. Note that in the pre-outburst data, background dominates above 2.5~keV; therefore, we only fit in the 0.7--2.5~keV range and due to the lower number of source counts we grouped that spectrum to have 50 counts per channel. We used the {\tt tbabs} model in Xspec \citep{Arnaud1996} assuming ISM abundances  \citep{Wilms2000}. Throughout the text, we present 68\% confidences for all the parameter values. To search for any statistically significant variation, we thawed all parameters of the model components and fit all spectra with only hydrogen column density being kept linked.  A blackbody plus a power-law model provides the best fit ($\chi^2$/dof = 1.07 for 2071 degrees of freedom, dof) with a resulting hydrogen column density $N_{H}$=(1.35$\pm$0.02)$\times$10$^{22}$~cm$^{-2}$. Note that a comparable fit can be obtained with a model comprising two blackbodies with a $\chi^2$/dof = 1.10 for the same dof, but with  a  smaller $N_{H}$, 0.90$\pm$0.01$\times$10$^{22}$~cm$^{-2}$.  For the post-outburst data, the resulting best-fit parameters and their time evolution are shown in Figure \ref{evolution} for the blackbody plus power-law model. Figure \ref{spectra} shows the extracted X-ray spectra together with this spectral model. 

We also extracted the \swift\ BAT survey data for the period MJD 58423--58524, when \xtej\ was observed with \nicer\ and was in the BAT FOV. The BAT data are processed via the {\tt batsurvey} pipeline script and with the off-axis correction. We do not find any significant emission in the BAT data, with 3$\sigma$ upper limits of $<$17 mCrab and $<$19 mCrab in the 14--24 and 24--50 keV bands, respectively. Thus, we do not add the hard X-ray component above 10 keV reported by \citet{Gotthelf2019} for the following NICER spectral analyses.

 The best-fit values for the pre-outburst data were $kT=0.112\pm0.003$~keV and $\Gamma$=4.92$\pm$0.25. The total absorbed flux in the 0.7--8.0 keV range is  $F_{\rm  tot}=(9.71\pm0.01)\times10^{-13}~{\rm erg}~{\rm s}^{-1}~{\rm cm}^{-2}$, in agreement with the quiescence flux level reported as 5--$10\times10^{-13}~{\rm erg}~{\rm s}^{-1}~{\rm cm}^{-2}$ \citep{Gotthelf2004,Ibrahim2004}. A two-blackbody model, for the pre-outburst data yields a $kT=0.41\pm0.06$~keV and $kT=0.161\pm0.004$~keV with a total absorbed flux of $F_{\rm  tot}=(9.49\pm0.04)\times10^{-13}~{\rm erg}~{\rm s}^{-1}~{\rm cm}^{-2}$ in the same range. We note that especially for energies above 5~keV calibration uncertainties do exist, which have the potential to bias the best fit power-law photon index. However, these effects should be minimal, given the fact that the results presented here are comparative and mostly related to lower energies.  

Modeling that assumes no spectral variation during the post-outburst interval does not result in an acceptable fit, with $\chi^2$/dof = 1.3 for 2099 dof. Still, as can be seen from Figure \ref{evolution}, beyond the first three segments, the inferred spectral parameters are all in very good agreement with each other. To better probe this trend we linked all the spectral parameters of the blackbody plus a power-law model for the first two segments and the last five segments. The third segment is excluded due to short on-source time resulting in large uncertainties in the parameters. This way we obtained a best fit with a $\chi^2$/dof = 1.089 for 2091 dof. The best fit blackbody temperature showed no significant variation between these two groups: kT=0.666$\pm$0.003~keV to kT=0.671$\pm$0.001~keV. The slope of the power-law, on the other hand, showed a marginal softening with best-fit values $\Gamma$=2.93$\pm$0.07 and $\Gamma$=3.09$\pm$0.06, respectively. The measured absorbed fluxes in the 0.7--8.0~keV range are $F_{\rm  tot}=(1.561\pm0.005)\times10^{-10}~{\rm erg}~{\rm s}^{-1}~{\rm cm}^{-2}$ and $F_{\rm  tot}=(1.466\pm0.003)\times10^{-10}~{\rm erg}~{\rm s}^{-1}~{\rm cm}^{-2}$, respectively, for the two groups of segments. In a similar manner, we see  marginal cooling of the  two blackbody components with temperatures changing from 0.48$\pm$0.04 and 0.75$\pm$0.03~keV to 0.38$\pm$0.09 and 0.71$^{+0.09}_{-0.04}$~keV, respectively. The two blackbody model results in a fit $\chi^2$/dof = 1.1179 for 2091 dof.

%%%%%%%%%%%%%%%%%%%%%%%%% Phase Resolved %%%%%%%%%%%%%%%%%%%%%%%%%%%%%%%%%%%%%%

\section{Phase Resolved Spectral Analysis}
\label{sec:phase}

As shown in the previous section, at least during the last five segments, the source flux and spectral parameters remained constant within errors. 
We, therefore, focused only on these five segments (MJD 58524$-$58529, with a total of 10.5~ks exposure) to explore the source's phase resolved behavior. Using the timing solution we obtained in Section \ref{sec:timing}, we extracted spin phase resolved X-ray spectra. As shown in Figure \ref{pf}, we divided the pulse profile into 4 intervals: $\phi$=0.92$-$1.12, 1.12$-$1.37, 1.37$-$1.52, and 1.52$-$1.92, as Minimum, Rise, Peak, and Decay, respectively. 

\begin{figure}[ht!]
\begin{center}
\includegraphics[scale=0.45,angle=0]{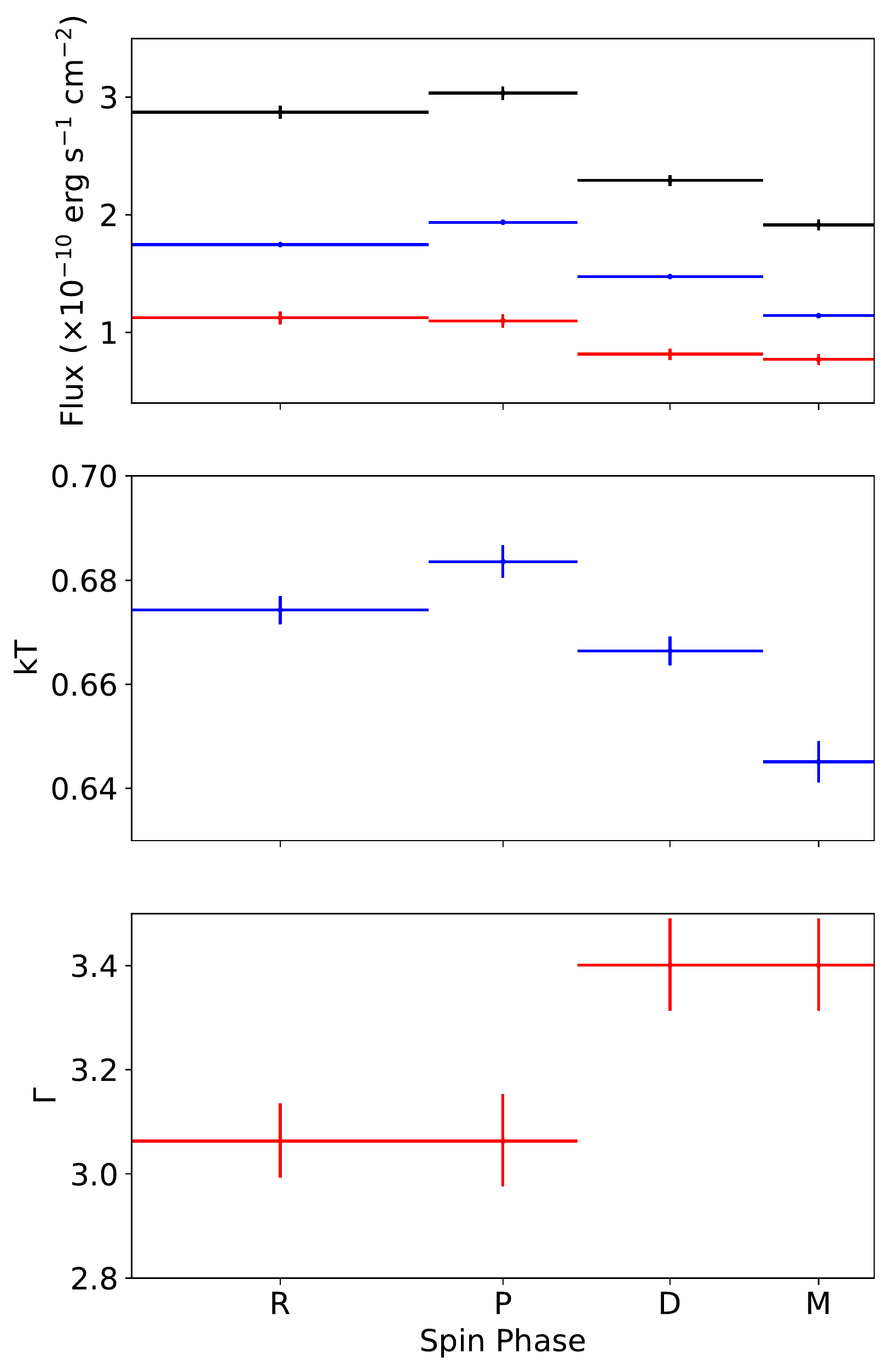}
\caption{Spin phase evolution of the blackbody plus power law model. The upper panel shows the unabsorbed fluxes of each component (thermal in blue, non-thermal in red) and the total flux (black) as observed in the 0.7-8.0~keV range. The middle and bottom panels show the phase evolution of the blackbody temperature and the photon-index. \label{phase_evolution}}
\end{center}
\end{figure}

Using the same approach as in Section \ref{sec:sp_analysis}, we grouped the X-ray spectra to have at least 200 counts per channel and fit each with the blackbody plus power-law model. Initially, we linked all the parameters between individual spin phase intervals, which resulted in unacceptable statistics with $\chi^2$/dof values of around 9.5, clearly indicating a spin phase dependence of the spectral parameters. We then allowed all of the parameters to vary between spin phases, again keeping $N_{H}$ linked. This way, it was possible to obtain a reasonable fit, with $\chi^2$/dof of 1.187/1120. The resulting $N_{H}$ was similar to the value obtained in Section \ref{sec:sp_analysis}.
Initial inspection of these results indicated that as the pulse decays and reaches the minimum, the X-ray spectrum significantly softens. Furthermore, the power law index of the rise and peak phase, as well as the decay and minimum phase are consistent within each other. When we link this parameter to have one value for the rise-peak and another value for the decay-minimum phases, the reduced $\chi^2$ slightly decreases to 1.185 for 1122 dof. In Table \ref{tab:phase_res} and Figure \ref{phase_evolution} we provide the results of this final fit. 

 \begin{table*}[t]
\begin{centering}
 \caption{Best fit parameters of the blackbody plus power-law model as a function of neutron star spin phase.}
 \label{tab:phase_res}
 %[loc]\scalebox{0.8}{
\setlength{\tabcolsep}{5.2pt}
\scriptsize{
 \begin{tabular}{cccccc}
  \hline
   Phase & Interval & kT & Flux$^{\dagger}$$_{BB}$ & $\Gamma$ & Flux$^{\star}$$_{PL}$  \\
 &  & (keV)  &    &  & \\
\hline 
Decay    & 1.12$-$1.37 & 0.666 ${\pm}$ 0.003 &  	 1.48 ${\pm}$ 0.02 &   3.40 ${\pm}$ 0.09 &  	 0.82 ${\pm}$ 0.05	 \\
Minimum    & 1.37$-$1.52 &	0.645 ${\pm}$ 0.004 &	 1.14 ${\pm}$ 0.02 &   --                &  	 0.77 ${\pm}$ 0.04	 \\
Rise   & 1.52$-$1.92 &	0.674 ${\pm}$ 0.003 &	 1.75 ${\pm}$ 0.02 &   3.06 ${\pm}$ 0.07 &  	 1.11 ${\pm}$ 0.06	 \\
Peak & 0.92$-$1.12 &	0.684 ${\pm}$ 0.003 &	 1.94 ${\pm}$ 0.02 &   --                &  	 1.09 ${\pm}$ 0.06	 \\
\hline 

\end{tabular}}
 \footnotesize{
   \begin{flushleft}
     $^{\dagger}$ Unabsorbed blackbody flux in the range of 0.7$-$8.0~keV in units of 10$^{-10}$ erg cm$^{-2}$ s$^{-1}$. \\ 
    $^{\star}$ Unabsorbed power-law flux in the range of 0.7$-$8.0~keV in units of 10$^{-10}$ erg cm$^{-2}$ s$^{-1}$. \\ 

\end{flushleft} }
 \end{centering}
\end{table*}

%%%%%%%%%%%%%%%%%%%%%%%%%%%%%%%%%%%%%%%%%%%%%%%%%%%%%%%%%%%%%%%%%%

\section{Discussion}
\label{sec:disc}

Our \nicer\ observations provide, for the first time, soft X-ray coverage of the early phase of an outburst from this unique magnetar. The soft X-ray spectrum is successfully fit with a black-body plus power-law model. Converted from the \swift\ BAT upper limits  (Section \ref{sec:sp_analysis}), the hard X-ray flux upper limit is estimated to be $2.6\times 10^{-10}$~ergs~s$^{-1}$~cm$^{-2}$ in the 15--60 keV band. 

We found that the source flux was 160 times higher than its pre-outburst level also measured with \nicer\ some 215 days before. We also found that, within the eight day span, the spectral parameters showed marginal variations. The 2--10~keV flux varied from 1.206${\pm}$0.007 to 1.125${\pm}$0.004 $\times10^{-10}~{\rm erg}~{\rm s}^{-1}~\rm{cm^{-2}}$ within our observing span. These flux measurements indicate a 44\% decline with respect to the {\it NuSTAR} measurement. A linear fit to the {\it NuSTAR} and two \nicer\ measurements yield a daily decline rate of 1.83$\times10^{-12}~{\rm erg}~{\rm s}^{-1}~\rm{cm^{-2}}$. However if we also take into account the earlier {\it MAXI} flux \citep{Mihara2018}, the decay is best approximated with an exponential function.

Our post 2018 outburst phase-connected solution spanning a time baseline of about eight days yields a spin down rate of 9$\times$10$^{-12}$ s~s$^{-1}$, which is almost 3.2 times larger than that reported by \citet{Pintore2019}. The magnetar has likely entered a higher $\dot{P}$ episode at the onset of the latest outburst. Variable spin-down in conjunction with outbursts is common among magnetars \citep{Woods1999, Gavriil2004, Kaspi2014, Archibald2015, Younes2015}. The extrapolation of the spin ephemeris of \citet{Pintore2019} to the epoch of our spin solution yields a spin frequency of 0.18045726~Hz, which is 1.3$\times$10$^{-7}$~Hz larger than our measurements. Even though it is possible to consider a timing (anti-)glitch at the onset of its latest outburst, the difference may have resulted from a non-secular spin down trend of the magnetar.

The energy dependence of the pulsed fraction and pulse phase is
consistent with that reported by \citet{Gotthelf2019}. The pulsed
fraction increases with energy up to 6.5~keV, and drops rapidly beyond
that. Similarly, pulse phase lags increase below 6.5~keV but
decrease at higher energies. Further investigations to address a possible connection between these two properties requires additional monitoring of the source.

Taking advantage of the source brightness, together with the large effective area of {\it NICER} in the soft X-ray band, we also performed a phase resolved spectral analysis by dividing the pulse profile into four intervals. We found a small but statistically significant variation in the blackbody temperature. Assuming a distance of 3.1~kpc, apparent emitting radius of the blackbody component varies from 2.4 to 2.8~km, between minimum and peak phases, respectively. These results on the thermal component indicate that the pulse period dependent flux modulation is mostly due to a hot spot on the surface. We also found that the power law index is steeper (at the 3$\sigma$ level) in the rise-peak interval than the decay-minimum one. Given that the power-law component is often attributed to the magnetospheric twist in magnetars \citep{Thompson2002,lyutikov2006,fernandez2007}, the change in index might indicate viewing of a slightly more twisted part of the magnetosphere during the decay-minimum phase. 

\acknowledgments
We thank Wynn C. G. Ho for careful reading and insightful comments on this manuscript.  We thank the anonymous referee for the careful reading of the manuscript and his/her thoughtful contributions. E. V. G. acknowledges NASA Grant 80NSSC18K0452. T. E. was supported by JSPS grant numbers 16H02198 and 18H01246. This work was supported by NASA through the {\it NICER} mission and the Astrophysics Explorers Program. Part of this research was carried out at the Jet Propulsion Laboratory, California Institute of Technology, under a contract with the National Aeronautics and Space Administration. This research has made use of data and software provided by the High Energy Astrophysics Science Archive Research Center (HEASARC), which is a service of the Astrophysics Science Division at NASA/GSFC and the High Energy Astrophysics Division of the Smithsonian Astrophysical Observatory.

\facilities{\textit{NICER}}

%%%%%%%%%%%%%%%%%%%%%%%%%%%%%%%%%%%%%%%%%%%%%%%%%%%%%



\end{document}